\documentclass[prd,twocolumn,superscriptaddress]{revtex4}

\usepackage{amsmath,amsthm,amssymb}
\usepackage{graphics,graphicx}
\usepackage{epsfig}
\usepackage{color}

\begin{document}
\title{Higher order corrections to Heterotic M-theory inflation}
\author{Paulo Vargas Moniz} \email{pmoniz@ubi.pt}\affiliation{Departamento de Fisica,
Universidade da Beira Interior, 6200 Covilha, Portugal.}
\affiliation{ CENTRA,  Instituto Superior T\'ecnico, 1049-001
Lisboa, Portugal}
\author{Sudhakar Panda} \email{panda@mri.ernet.in}\affiliation{Harish-Chandra Research Institute, Allahabad-211019, India.}
\author{John Ward} \email{jwa@uvic.ca}\affiliation{Department of Physics and Astronomy, University of Victoria, Victoria, BC, V8P 5C2, Canada.}
\begin{abstract}
We investigate inflation driven by $N$ dynamical five-branes in Heterotic
 M-theory using the scalar potential derived from the open membrane instanton sector. At leading order the resulting theory can be mapped
to power law inflation, however more
generally one may expect higher order corrections to be important. We consider
a simple class of such corrections, which imposes tight bounds on the
number of branes required for inflation.
\end{abstract}
\maketitle

\section{Introduction}
There has been a significant effort
in recent years to better
understand  moduli stabilisation and
cosmological dynamics in a specific class of string theory models.
More precisely, whilst much of this
effort has been focused on the type IIB superstring, there has been
significant progress in the
strongly coupled Heterotic models which are conjectured to be the
low energy limit of Heterotic M-theory \cite{Horava:1995qa,
Horava:1996ma}. These models are phenomenologically interesting because of the $E_8 \times E_8$ gauge
group which comprises a visible sector (including the standard model) and hidden sector.
The embedding of wrapped five-branes in this
setting was investigated and
suggested as an ideal method of driving
inflationary cosmological scenarios
\cite{Bailin:1998tu,Lukas:1999kt,Lukas:1998sy,Gray:2003vk,
Copeland:2001zp}. The fact that it
naturally contains a dark sector, which is coupled gravitationally
to the visible sector, suggests a natural place to localise dark
matter making such  models
phenomenologically relevant. With the advent of flux
compactification, various groups tried to develop more realistic
models of the vacuum structure of the theory \cite{Becker:2003yv,
Curio:2001qi,Buchbinder:2003pi,Becker:2004gw} which led to the
embedding of a standard model-like sector \cite{Braun:2005fk} into
the Heterotic theory using the vector-bundle moduli.

Inflationary model building in the
Heterotic model \cite{Buchbinder:2004nt, deCarlos:2004yv} was
supported by precision data retrieved from WMAP. This  culminated
in the multi-brane model of \cite{Becker:2005sg} (supplemented by
follow-up work in \cite{Krause:2007jr, Ward:2005ti}) and other
related ideas \cite{Ashoorioon:2006wc}. The important point about
\cite{Becker:2005sg} is that it provided a concrete embedding of
assisted inflation \cite{Liddle:1998jc,
Kanti:1999vt,Kanti:1999ie,Copeland:1999cs} into a fully UV-complete
theory, where the inflationary phase occurs \emph{before} the end of moduli
stabilisation. Assisted inflation works for theories with steep
scalar potentials, provided that there are multiple scalar fields
following similar trajectories through field-space. The combined
effect of the multiple fields acts
to enhance the Hubble friction term
in the field equations, thus allowing sufficient inflation to occur.
In this framework, the scalar
potential can be significantly
steep to ensure that the dynamics of the geometric moduli can be
ignored, therefore indicating that
assisted inflation would be the most
likely possibility for driving inflation in this model. We also note that the end of
inflation is naturally understood to occur through the tunneling of
branes into the orbifold planes via instanton transition into the
boundary, see \cite{Battefeld:2008py, Battefeld:2008qg} for some
related work. After inflation, the remaining moduli can be
stabilised in a de-Sitter vacuum \cite{Buchbinder:2004im,
Hsu:2005ke, Curio:2006dc} which is well controlled due to the known
higher derivative corrections \cite{Anguelova:2005jr}. Unfortunately
stabilising the more general $SU(3)$ structure case is still a work
in progress, although it is likely that including gaugino
condensation and following the proposals \cite{Anguelova:2006qf,Gray:2007zza} will achieve this.

The model of assisted inflation arises very naturally in this
instance by considering only the leading order terms in the
superpotential \cite{Becker:2005sg}. A natural
question to ask then relates to whether inflation is spoilt through
the inclusion of higher order terms, since this would indicate
another (potentially un-natural) source of fine tuning. We aim to go
beyond the $1/N$ expansion by considering additional corrections to
the tree-level superpotential, as loop corrections are difficult to
calculate explicitly, to further explore the allowed inflationary
parameter space. Our naive expectation is that these corrections
will impose tighter constraints on the background parameters in
order for inflation to occur - however since the scalar potential is
comprised solely of exponential terms we anticipate that inflation
(if it occurs) could be eternal. To check this assumption it is
useful to study the phase space in order to find attractor
solutions.

The organization of this note is as follows. In section
\ref{background} we introduce the relevant M-theory background
required for model building. We follow this in section
\ref{coldinflation} by discussing assisted inflation, and how it is
realized in this theory. In sections \ref{newsec-1}, \ref{newsec-2}
and \ref{newsec-3} we explore and explain how this setting can be
made more accurate through {several
new}  additional elements. The main contribution of this paper is in
providing and discussing
the cosmological implications of
these elements, which are higher
order corrections to the inflationary
potential, and discuss
how they further restrict the inflationary parameter space.


\section{The M-theory background}\label{background}
Let us consider the model of the strongly coupled Heterotic string
on the orbifold $S^1/ \mathbb{Z}_2$, where there are two 'end of the
world' branes localised at the fixed points of the orbifold action
\cite{Horava:1995qa, Horava:1996ma}. For simplicity we will
consider a simplified model, whereby we
restrict ourselves to a sector where $h^{1,1}=1$, and where there
are $N$ five-branes wrapping an isolated genus-zero curve in the
$CY_3$, localised along the orbifold direction. The superpotential
for the theory takes the following form \cite{Becker:2005sg}
\begin{equation}
W = W_{flux} + W_{OM}- W_{GC}
\end{equation}
where we have defined each of the contributions to be
\begin{eqnarray}\label{eq:superpotential}
W_{flux} &\sim & \int H \wedge \Omega \nonumber \\
W_{OM} & \sim & h \left(e^{-T}+\sum_{i=1}^N e^{-Y_i}+\sum_{i=1}^N e^{-(T-Y_i)} \right) \nonumber \\
&+& h \sum_{i<j} e^{-(Y_j-Y_i)} \\
W_{GC} &\sim& C_H \mu^3 \exp \left(-\frac{1}{C_H} \left \lbrack S+\gamma_h T + \sum_{i=1}^N \frac{\gamma Y_i^2}{T}\right\rbrack \right) \nonumber
\end{eqnarray}
in terms of $\mathcal{N}=1$ chiral superfields. Note that $\mu$ is a parameter fixed by the UV-cutoff of the gauge group of the hidden theory, $\gamma$ is
the normalised coupling of the five-branes, whilst $\gamma_h$
is given by the normalised integral of the K\"ahler form over the relevant two-cycle.
Let us briefly comment on some of the notation
here, since it is different to that employed in the type II case.
Firstly, it is the superfield $S$ that controls the overall volume
$\mathcal V$ of the compact $CY_3$. The superfield $T$ measures the
length of the orbifold, and $Y$ is the superfield associated with
the location of the fivebrane along this interval. Strictly speaking
each of these superfields contains axionic contributions associated
with the reduction of the three-form flux along the orbifold
direction
\begin{eqnarray}
S &=& \mathcal{V} + \mathcal{V}_{OM} \sum_{i=1}^N \left(\frac{x_i}{L} \right)^2 + i \sigma_s \label{3a} \\
T &=& \mathcal{V}_{OM} + i \sigma_T  \label{3b}\\
Y_i &=& \mathcal{V}_{OM} \left(\frac{x_i}{L} \right) + i\sigma_i.
\label{3c}
\end{eqnarray}
Clearly $x_i$ measures the distance of the $i$th brane from the
visible sector, therefore $x_i \in [0,L]$. We also have
contributions from open-membranes wrapping cycles within the $CY_3$,
which have an averaged volume given by $\mathcal{V}_{OM}$. These
open-membrane contributions can be interpreted as a shift in the
volume of the internal space and have phenomenological values
$\mathcal{V}_{OM} \sim \mathcal{O}(1)$, whilst typically $\mathcal{V} \sim \mathcal{O}(10^2)$.
We will often refer to 'canonical values' for these parameters which are $\mathcal{V}_{OM} \sim 7, \mathcal{V} \sim 341$
as an approximate solution \cite{Becker:2005sg}, and therefore we can,
in principle, write the superfields solely as functions of $N$. Of course
if the hidden sector gauge group is no longer $E_8$, then the dual Coxeter number is different and one obtains
different values for these parameters, eg $\mathcal{V}_{OM} \sim 21$, $\mathcal{V} \sim 215$ if the group is $SO(10)$
\cite{Becker:2004gw}. Also note that $\sigma$ refers to the axionic component of each of the superfields.

In the flux superpotential (\ref{eq:superpotential}) $H$ is the
three-form flux which is to be integrated over the manifold, and
$\Omega$ is the usual $(3,0)$ form.
The presence of this term is crucial in order to stabilise the
complex structure of the Calabi-Yau manifold. The gaugino condensate
contribution ($W_{GC}$) contains various parameters proportional to
the Coxeter number $C_H$ of the condensing gauge group - and various
other integer terms parameterised by the $\gamma$ functions which
arise from integrating the Kahler form over the corresponding
internal cycles. As in the case of type II string theory, this term
plays a vital role in the stabilisation of the compact space.

The open membrane contribution ($W_{OM}$) arises from Euclidean
membranes fully wrapping cycles within the $CY_3$. The various terms
correspond to instantons between the visible sector and the brane,
the brane and the hidden sector and the visible and hidden sectors
respectively. The prefactor $h$ is the Pfaffian associated with the
hidden sector. It is typically a function of the vector bundle
moduli living on the hidden sector domain wall - however we will
ignore these in our toy model construction, although they play a
vital role in realising models of the MSSM. It is this
superpotential contribution which will be the relevant one for
inflationary model building.

Having specified the superpotential, we must also consider the Kahler potential for the theory which (as in the type II case) is separable at tree level
\begin{eqnarray}\label{eq:kahler_potential}
K &=& -\ln\left(S+\bar{S} + Z - \sum_{i=1}^N \frac{(Y_i+\bar{Y_i})^2}{T+\bar{T}} \right) - 3 \ln (T+\bar{T}) \nonumber \\
&+& K_{CS}
\end{eqnarray}
where $K_{CS}$ corresponds to the potential on the complex structure moduli space. It will usually be simpler to introduce the function
\begin{equation}
Q = S+\bar{S} +Z - \frac{(Y+\bar{Y})^2}{T+\bar{T}}
\end{equation}
to reduce the expressions above. For simplicity we will stabilise
the complex structure at high scales allowing us to integrate them
out of the theory. Also note that (for completeness) we have
included the presence of the leading order correction to the theory
coming from the $R^4$ terms in the eleven dimensional action,
denoted by $Z$, which is a function of the Euler number of the
$CY_3$ and is therefore a topological restriction. We will set it to
be zero for the time being.

The second term in (\ref{eq:kahler_potential}) should be familiar from the type II theory, and is explicitly a function of the Kahler moduli only. Since
we have assumed that there is only a single modulus in this sector, the intersection numbers become trivial. Extending this to the multi-moduli
case is technically more complicated, although can be done in principle.

\section{ Fivebrane inflation}\label{coldinflation}

Having set up the relevant background, one can now describe how
fivebrane inflation works in this set-up \cite{Buchbinder:2004nt,
deCarlos:2004yv, Becker:2005sg}. Modular inflation driven by either
the geometric moduli themselves \cite{Banks:1995dp, Binetruy:1986ss, Conlon:2005jm, Bond:2006nc}, or their axionic partners has previously been
considered in the literature \cite{Grimm:2007hs, Kallosh:2007cc, Misra:2007cq}.
In this paper we are interested in inflation
driven by the dynamics of five-branes along the orbifold length,
which is the M-theory counterpart of $D$-brane inflation
\cite{Dvali:1998pa}. In order to proceed, one must have parameteric
control over the theory. In particular this usually means
stabilising all the moduli of the compactification prior to the
inflationary phase. This is still a non-trivial task in the
heterotic context. Fortunately the presence of multiple branes
wrapping genus zero curves, allows us to by-pass this condition (at
least for the purposes of inflationary model building). It turns out
that the scalar potential is actually steepest along the five-brane
directions, which leads to the possibility of inflation occuring
before the stabilisation of the (pseudo)moduli - where the inflaton
direction is identified with the steepest part of the potential
\cite{Becker:2005sg}.

Standard slow roll inflation relies on the fact that the inflaton
potential is extremely flat, therefore the acceleration of the
inflaton field can be neglected - at least for early times - and the
universe will undergo sufficient expansion. However a scalar
potential that is exponentially decreasing is always too steep for a
sustained period of cosmological inflation in this context. The
inflaton undergoes rapid acceleration and the slow roll
approximation rapidly breaks down. One simple way to alleviate this
problem is to consider not just a single field for the inflaton, but
several fields that all have the same trajectory. The combined
effect of these $N$ fields is to increase the strength of the Hubble
term in the inflaton equation of motion, which acts much like a
friction term. The more fields we can include, the more the motion
will be damped because the friction is increasing. This was given
the name assisted inflation in the literature.

The model of assisted inflation we describe herein, actually belongs
to a special class of models which overlap between assisted
inflation and power law inflation - so called because the scale
factor expands like $a(t) \sim a_0 t^p$. In terms of an effective
description, it is more useful to parameterize power law inflation
through the introduction of an exponential potential of the form
\begin{equation}
U(\phi) \sim U_0 e^{\phi/\sqrt{p}}
\end{equation}
(where we work in Planckian units) with $\phi$ corresponding to a
canonical scalar field. The resulting slow roll parameters can be
seen to reduce to constants. We find that $\epsilon = 1/p$, and
therefore since inflation demands $\epsilon << 1$ we simply need to
ensure that $p>>1$ for inflation to occur. One potential difficulty
with these solutions is that there is no natural exit from
inflation, since the relevant parameters are constants - therefore
some other mechanism must be invoked to end power law inflation.

Returning to the case at hand, we include $N$ five-branes, wrapping
two-cycles within the compact space which are constrained to move
along the orbifold direction, and fill the large $3+1$ dimensional
space-time. Let us briefly discuss some of the technical aspects
associated with such a construction. Firstly we will take the
Horava-Witten model, which assumes that the seven-dimensional
compact space is simply the product orbifold $CY_3 \times S^1/Z_2$,
where the $CY_3$ is a manifold with $SU(3)$ holonomy. More generally
one should consider compactifications on full $G_2$ holonomy
manifolds, or $X_6 \times S^1/Z_2$ orbifolds where $X_6$ is a
manifold of $SU(3)$ structure. Thus our choice of background is
already subject to some tuning of the initial conditions, in that we
are selecting a specific compactification manifold from the space of
all such manifolds. It may well turn out that the $CY_3$
compactification is the more dynamically favoured one. However in
the absence of a fuller understanding of M-theory we must include
this as part of our initial conditions. Having specified our compact
space we must then be careful to wrap the five-branes only along
genus zero curves, since terms arising from branes wrapped on higher
genus curves will vanish due to holomorphy and therefore will not
contribute to the theory. Finally we will assume that the fivebranes
wrap these cycles only once - as this simplifies things considerably
since the anomaly cancellation
expression then reduces to
\begin{equation}
\beta_v + \beta_h + N = 0
\end{equation}
where $\beta_{v/h}$ are integer coefficients arising from the second
Chern-Classes on the visible/hidden
boundary respectively.

Given these constraints we find that the superpotential is dominated by terms coming from the open-membrane instantons
\begin{equation}
W_{OM} \sim h\left( e^{-T} + \sum_{i=1}^N e^{-Y_i} + \sum_{i=1}^N e^{-(T-Y_i)}+ \sum_{i<j} e^{-(Y_j-Y_i)} \right)
\end{equation}
with the Kahler potential given as in
(\ref{eq:kahler_potential}). It was then argued in
\cite{Becker:2005sg} that the initial conditions require the
five-branes to be localised in the middle of the orbifold, where the
total orbifold length is much larger than the other scales in the
theory. This ensured that the first three terms in the
superpotential can be neglected as being exponentially suppressed.
This then leaves only the term arising from interactions between
each of the five-branes
\begin{equation}
W_{55} \sim h \sum_{i<j} e^{-(Y_j-Y_i)}.
\end{equation}
It is a standard result in $\mathcal{N}=1$ supergravity that the
scalar potential is minimised at supersymmetric vacua where
$D_{\alpha}W=0$, with $\alpha$ running over the various moduli in
the theory. Therefore it is energetically favourable for us to set
$D_{Y_i}W=0$. With this assumption we find the constraint
\begin{equation}
W_{55} \sim e^{-K}
\end{equation}
where $K$ is the full Kahler potential. Neglecting the complex
structure term then amounts to trying to solve the expression
\begin{equation}
\sum_{i<j} e^{-(Y_j-Y_i)} + (T+\bar{T})^2 \sum_{i=1}^N (Y_i+\bar{Y_i})^2 = (S+\bar{S}+Z)(T+\bar{T})^3
\end{equation}
which is generally very difficult to do analytically.

Consequently in \cite{Becker:2005sg} they opt to take a
simpler route, namely to individually set each term in the
covariant derivative to zero - i.e. they impose
\begin{equation}
\partial_{Y_i}W_{55} = W_{55} \partial_{Y_i}K = 0.
\end{equation}
In more detail this can be summarised as:
\begin{itemize}

\item  The first constraint basically tells us that the
superpotential is independent of each of the individual superfields.
The only way for this to be non-trivially satisfied (i.e not having
a vanishing superpotential) is for $(Y_j - Y_i)$ to be a constant. This is
therefore a geometrical condition since it implies that the
five-branes should be equidistant from each other.

\item The secondary constraint is that $K_i = 0$ which reduces to
\begin{equation}
\frac{2}{Q}\frac{1}{(T+\bar{T})} \sum_{i=1}^N (Y_i + \bar{Y}_i)^2 = 0
\end{equation}
which is assumed to be true for sufficiently large values of $Q(T+\bar{T})$.

\item The final assumption was to keep only
nearest-neighbour instantons contribute to the superpotential.
Defining the quantity $\Delta Y = Y_{i+1}-Y_i$ one sees that the
superpotential reduces to
\begin{equation}
W_{55} \sim h \sum_{i=1}^{N-1} e^{-\Delta Y} = h (N-1) e^{-\Delta Y}
\end{equation}
which scales linearly with $N$, and therefore is the leading term in a $1/N$ expansion.
\end{itemize}

Given the above assumptions the F-term scalar potential can be seen to schematically reduce (at leading
order) to the form
\begin{equation}
U_F \sim U_0 e^K |W_{55}|^2 \sim {\rm const} |W(Y_i)|^2
\end{equation}
where the superpotential term is a strictly decaying exponential
contribution.
This exponential suppression of the inflaton
{constitutes} the dominant mechanism
for building concrete inflationary models within superstring
theory\footnote{For example in the context of type II theory we see
that the inflaton potential is of the form $U_F \sim U_0 \left(1-h
e^{-\tau}+\ldots \right)$, where $\tau$ is related to one of the
Kahler moduli in the large-volume models in IIB, or one of the
complex structure moduli in the weakly-coupled models of type IIA.
The main difference between these approaches and the one in the
heterotic theory are to do with moduli stabilisation. The models of
the type II theory have all their (heavy geometric) moduli stabilised
initially at some high scale allowing inflation to occur through the
dynamics of the lightest moduli, whilst in the heterotic M-theory
model; none of the moduli are stabilised prior to inflation. In the
type II context, most of the moduli have already been stabilised
assuming zero temperature and therefore simply turning on finite
temperature effects may not be completely consistent since this may have the effect
of shifting the vev's of the fields - potentially destroying the
nice inflationary and phenomenological properties of such models.}.

Now since the branes are assumed to be equidistant, one can use their combined
centre of mass to re-write the superpotential in terms of a
canonical field $\psi$. In terms of which we see that the
superpotential becomes
\begin{equation}
W_{55} \sim (N-1) h e^{- A \psi/2}
\end{equation}
where $A$ is a function of the other geometric moduli $(s, t)$ via
$A \sim \sqrt{12 s t/(N(N^2-1))}$.
Note that our conventions differ from that in \cite{Becker:2005sg} thus accounting
for the change in the definition of the $A$ parameter, but $(s,t)$
are still associated with the real parts of the superfields
as in eqn (\ref{3a}-\ref{3c}). Using the
definition of the scalar potential above we see that the inflaton F-term
potential for $\psi$ is in fact
\begin{equation}\label{eq:powerlawfterm}
U_F \sim U_0 \frac{(N-1)^2}{s t^3} e^{-A \psi}
\end{equation}
where it was assumed that $(s,t)$
are approximately constant during inflation. One can check that the
potential along the inflaton direction is still actually the
steepest potential in field space, despite the apparent $1/t^3$
dependence on the orbifold modulus.

It is straight-forward to see that this exponential potential can
then be mapped to that of the power law form and we then obtain
power law inflation. In this case there is also a natural mechanism
for the end of inflation, since eventually the orbifold length $t$
will grow larger and then induce gaugino condensation on the hidden
boundary. The remaining terms in the superpotential will then be
non-negligible and also contribute to the F-term scalar potential.
The combination of higher order instanton effects and gaugino
condensation should then stabilise all the remaining geometric
moduli. Before proceeding, let us
stress that the above model works precisely because we assume the
geometric moduli can be taken to be approximately constant during the
inflationary phase driven by the exponentially decaying inflaton. In addition
the assumption that \emph{all} the branes are localised near $x \sim L/2$ is 
crucial in obtaining the simple form for the scalar potential

In the remainder of this paper we will describe how the setting
above can be further elaborated by means of
{additional  elements present in the
theory, which have not yet been fully explored.}
\section{Constraining $N$} \label{newsec-1}

Let us {appraise}  how all the
assumptions above affect the bounds on the number of branes. One of
the interesting features of this model is that $N$ is bounded from
both above and below \cite{Becker:2005sg} if one searches for
inflating trajectories. Indeed at leading order the scalar index is
related to the number of branes and therefore one can use the WMAP
normalisation to cherry-pick the requisite solution. Rather than
consider the inflationary normalisation, let us
{instead} see how the causal
structure of the theory imposes constraints on $N$. Recall that a
major simplifying assumption in the derivation of the F-term
potential (\ref{eq:powerlawfterm})
 was that we assumed $Qt >> 2 y^2$, which is a significantly stronger bound than simply assuming $Q>0$
which comes from the reality of the Kahler potential.
{We will}  be more careful than this
and examine the full constraint here.

Firstly let us consider the situation at the very start
of inflation. The initial distribution of branes is assumed to be
localised around the middle of the orbifold, therefore we have
$x^i/L \sim 1/2$ and thus $y^i \sim t/2$ once we set the axions all
to zero. Since we must demand that $Q > 0$ for the Kahler potential
to be well defined, this immediately yields the initial
upper bound that
\begin{equation}
N < \frac{2(2s+Z)}{t}
\end{equation}
where we have included the term coming from the $R^4$ corrections for generality.
Setting $Z=0$ initially, and fixing $s\sim \mathcal{O}( 10^2)$ and $t\sim
\mathcal{O}(1)$ to be constant - which we assume will be typical
values of the parameters at this time - we see that $N <
\mathcal{O}(100)$ as an order of magnitude approximation. However if we wish
to consider a more realistic bound then we must recall that $s$ is
also a linear function of $N$, and therefore does not provide a useful bound on $N$.
We can however bound the size of the correction term
$Z$, which must satisfy
\begin{equation}
Z > -2 \mathcal{V}
\end{equation}
where we remind the reader that $\mathcal{V}$ is the averaged volume of the Calabi-Yau. Since $Z$ is a function of the Euler number of the particular manifold
this in turn imposes a constraint on the number of complex structure and Kahler moduli i.e this is a geometric condition.

Regarding the inflationary bound on $N$, recall that for power law inflation to work,
we must decouple the higher order $y$ dependence from the scalar potential which requires
the following condition to be satisfied (from a detailed analysis of the scalar potential)
\begin{equation}
6t^2 Q^2\left((2s+Z)t-6y^2 \right) + 64 y^6 >>0.
\end{equation}
If we make the usual assumption regarding inflation occuring when the branes are near the middle of the orbifold then we can rewrite this
constraint as
\begin{equation}
8\mathcal{V}^3 + Z^3 + 6\mathcal{V}Z(\mathcal{V}+Z)-N\mathcal{V}_{OM}(2\mathcal{V}+Z)^2+\frac{N^3 \mathcal{V}_{OM}^3}{6} >> 0
\end{equation}
As far as phenomenological implications go, the constraint equation above can be interpreted as constraining the $\lbrace N, Z \rbrace$ solution space
for fixed volumes. The corresponding phase space can be seen to be concave near $Z \sim 0$ indicating that the constraint is
more easily satisfied for larger values of $Z$ (for a given choice of volume).

The reality condition on the parameter $Q$ can then be imposed. Since there is no problem for $Z>0$ we will only focus on compactifications which
have $Z\sim0$ or $Z \to - 2\mathcal{V}$ as these are the two most interesting cases.

Let us begin with the former condition, where we find the constraint
equation can be written as $8 \mathcal{V}^3 \alpha >> 0$
and demanding positivity of $\alpha$ is equivalent to the condition
\begin{equation}\label{eq:alpha_condition}
\frac{B}{2} \left(1-\frac{B^2}{24} \right) < 1, \hspace{0.5cm} B = \frac{N\mathcal{V}_{OM}}{\mathcal{V}}.
\end{equation}
Since we must assume the validity of the supergravity approximation we require large volumes, and therefore
the constraint is satisfied for all values of $B$ despite the term on the left hand side
not being a monotonic function.

If one assumes that $\mathcal{V}$ is sufficiently large, then the leading term in (\ref{eq:alpha_condition}) imposes a bound on $N$
\begin{equation}\label{eq:Nbound}
N < \frac{2 \mathcal{V}}{\mathcal{V}_{OM}}
\end{equation}
thus providing a constraint on the number of five-branes. In fact
this full bound is a little tighter than the one considered in \cite{Becker:2005sg}
using the standardised assumption that $\mathcal{V} \sim
341$ and $\mathcal{V}_{OM} \sim 7$ since we find that $N$ is
constrained viz $N < 98$.

Now let us consider the limit $Z \to -2 \mathcal{V}$, where the constraint equation simplifies considerably.
In fact the relevant positivity condition reduce to $ 1 + B^3/72 >0$
which is satisfied for all values of $B$ and therefore does not constrain the number of five-branes. Of course one must be careful with the physical interpretation
of this result because the Kahler potential is actually no longer well defined in this limit.

Another  immediate limitation of the
setting in \cite{Becker:2005sg} is related to the vanishing of the
five-brane F-terms. We recall that this meant setting
$\partial_{Y_i}W$ \emph{and}
$\partial_{Y_i}K \sim 0$ in the covariant derivative. The first
constraint is actually remarkably robust and would seem to be the
simplest choice possible. However
the second constraint requires the first derivative of the Kahler
potential along the $Y_i$ direction to vanish at large volume.
Whilst this is an adequate assumption, one must be careful to check
that this is consistent with the functional form of the scalar
potential. In other words, if we drop these terms, then we cannot
keep terms of the same order arising from the other F-terms. Indeed
all of the (first) Kahler derivatives will have terms of order $1/Q$
and so $Q>>1$ on its own is not a consistent approximation.

Relaxing this assumption makes things slightly more complicated as there are now off-diagonal pieces contributing to the inverse Kahler metric on field space.
Fortunately all the F-terms are independent of the superpotential, and therefore one only needs to study the Kahler potential. After a laborious calculation,
and assuming that $Q$ is dominated by the volume factor as before, we find that the relevant constraint to decouple the additional $y$ dependence becomes
\begin{equation}
y^2 << \frac{t(2s+Z)(5N-6)}{8(N-2)(2s+Z)-(5N-6)}.
\end{equation}
which can easily be satisfied for a range of $s,t$. Note that this is a weaker bound than the one obtained by studying the decoupled inflaton mode and
therefore we will not study it further.
\section{Dissolving branes}
\label{newsec-2}
The results presented in
\cite{Becker:2005sg} can be further re-interpreted as follows.  An
interesting issue, not yet fully explored, relates to the growing
size of the orbifold. Recall that the basic theory of instanton
inflation here relies on the fact that inflation happens rapidly,
well before the five-branes move far from their initial positions.
It is expected that power law inflation will naturally
self-terminate once the (initially) subleading instanton terms
become comparable to those
associated with the inflaton sector. Therefore in order to study the
end of inflation, one must also include these contributions. This is
technically a difficult problem. The fivebranes will eventually
dissolve into the boundary branes through small instanton
transitions, as discussed in \cite{Becker:2005sg}. This was
discussed in \cite{Battefeld:2008py}, where it was found that
inflation ended rapidly upon the inclusion of a $\dot{N}/N$ term.
However with regard to the above discussion the analysis in this
instance is not strictly correct. The reason is that the effective
potential (\ref{eq:powerlawfterm}) is no longer valid
once the five-branes are near the boundaries. As we have shown, the
model of power law inflation is \emph{only} valid for
nearest-neighbour interactions and \emph{only} when all the branes
are localised near the center of the oribifold. Once the branes move
a distance away from the centre point, the inter-brane
superpotential becomes of similar magnitude to the other terms in
the open-membrane superpotential and these terms must also therefore
be included in the analysis.

Let us assume that the first brane (i.e the brane closest to the
visible sector at smallest $x_{1}/L$) is near the visible boundary
at some reference distance $x_1 \sim \delta$ where $\delta$ is
assumed to be small and tending toward zero since we are taking
$x_{1} << L$. We will keep the approximations that the five-branes
are equidistant and also that only NN (Nearest-Neighbour)
instantons are important. Since the branes
are spread over the entire orbifold one sees that the cumulative
effect of the inter-brane instantons is actually suppressed. Now in
terms of the superfields we again demand that the branes are
equidistant and interact only with their nearest neighbours. This
means we can again write $Y_{j+1} - Y_j = \Delta Y$. Now it is
straightforward to see that the $i$th brane is located at $Y_i = Y_1
+ (i-1)\Delta Y$ from the visible sector, where $Y_1 = \alpha T$ for
simplicity.

One can then see that $e^{-Y_i}$ and $e^{-(T-Y_i)}$ are actually the
same in this limit so the contribution to the superpotential
simplifies. This is really just an artifact of the $\mathbb{Z}_2$
symmetry due to the orbifolding. Since we have singled out $Y_1$ as
being special, our summations now run over the other $N-1$ branes
giving a superpotential of the form
\begin{eqnarray}
W &\sim& 2h\sum_{i=2}^N e^{-Y_i} + he^{-T}+h(N-1)e^{-\Delta Y} \nonumber \\
&\sim& \frac{2h e^{-\alpha T}}{(1-e^{\Delta Y})} \left(e^{-(N-1)\Delta Y}-1 \right) \nonumber \\
&+& he^{-T}+h(N-1)e^{-\Delta Y},
\end{eqnarray}
where we have included the contributions from all the five-branes in
the stack. Since $T$ is growing at this point, the gauge coupling on
the hidden sector is increasing and generates a non-zero term coming
from gaugino condensation. This term can be computed to yield
\begin{eqnarray}
W_{GC} &\sim& C_H \mu^3 \exp^{-f/C_H} \\
f&=& S + \gamma_h T + \frac{\gamma}{T} \left(\alpha^2 T^2(N-1)+\alpha T \Delta y(N^2-1) \right) \nonumber \\
&+& \frac{\gamma}{T} \frac{N \Delta Y^2}{6}(1+2N^2-3N)
\end{eqnarray}
which one can see will lead to a significantly complicated expression for the F-term scalar potential.
\section{Inflation beyond the leading order approximation}
\label{newsec-3}

{Finally, in this section, we discuss
a } simple limitation with the power law
model{ which is related} to
the functional form of the superpotential. Indeed for inflation to
occur in this model without fixing the geometric moduli, we must
tune the superpotential in two ways. Firstly we assume that the $N$
branes are distributed over some length scale $\delta x^{11}$ which
is much smaller than the orbifold length, i.e $\delta x^{11} << L$.
This ensures that we can neglect the superpotential terms coming
from brane-boundary and boundary-boundary instantons. However an
immediate collorary is that the $N$ branes must be relatively close
together, and therefore one would assume that there can be a
sizeable contribution from \emph{all} the inter-brane instantons and not
just the nearest neighbour interactions (which was a crucial
assumption in the derivation of the power law potential). Let us
calculate the corrections to the potential arising from the latter
of these two effects, namely let us relax the condition that only
nearest neighbour (NN) interactions occur.

Assuming that the branes remain equidistant, the corresponding
contribution to the superpotential can then be written as
\begin{equation}
W_{55} \sim h \sum_{a=1}^{X} \sum_{i=1}^{N-a} e^{-a  \tilde{A} \psi}
\end{equation}
where $X$ denotes the instanton corresponding to the $X$th nearest
neighbour interaction and $\tilde{A}$ is defined as before but
includes the additional factor of two for simplicity. Note that this
is the generalisation of the expression first obtained in
\cite{Ward:2005ti}{.} Explicitly
performing the sum above yields the following F-term potential:
\begin{eqnarray}
U_X &\sim& \tilde U_0 h^2 e^{-2 \tilde A \psi} g(\psi) \\
g(\psi) &=& \left (\sum_{k=1}^X (N-k)e^{-(k-1)\tilde A \psi} \nonumber \right)^2,
\end{eqnarray}
where we have absorbed the factors of $(s,t)$ into the definition of
$\tilde U$ which we again take to be constant during inflation. One
can check that the above potential reproduces the leading order
solution {(cf. \cite{Becker:2005sg})}
in the limit that $X\to1$. Therefore corrections due to $X
> 1$ correspond to deviations from the power law inflationary
behaviour {(cf. Sect.
\ref{coldinflation}).}

The combined effect of the higher order instanton contributions clearly breaks the power law behaviour. Physically we may anticipate that the higher order
(large $X$) interactions may well be negligible, however the NNN (Next to Nearest Neighbour) interactions may well be important.
For example we see that if we keep only next to nearest neighbouring instanton terms then the potential reduces to the following
\begin{equation}
U_2 \sim \tilde{U_0} h^2 e^{-2 \tilde{A}\psi} \left((N-1) + (N-2)e^{-\tilde{A}\psi} \right)^2.
\end{equation}
The NNN correction leads to an expansion in powers of $1/N$. For small $\tilde{A}$, which we assume to be phenomenologcally favoured, this decouples the inflaton
at leading order allowing us to expand the first slow roll parameter $\epsilon_2$ as
\begin{equation}\label{eq:epsilon2}
\epsilon_2 \sim \frac{9 \tilde A^2}{2} \left(1-\frac{1}{3N} + \ldots \right)
\end{equation}
which is significantly larger than the NN case (which scales like $2 \tilde{A}^2$ in the notation of this section) - although still corresponding to
a power law scenario.
The result is intuitively obvious, namely that inflation is highly sensitive to the number of branes in the theory.

Since inflation must occur whilst the branes are localised near the center of the orbifold, corresponding to $\psi \sim 0$ in this notation, we see that the scalar potential
near the start of inflation reduces to
\begin{eqnarray}
& &U_X (\psi \sim 0) \sim \frac{X^2 U_0 h^2}{4} (2N-1-X)^2 - \nonumber \\
& &\frac{X^2 U_0 h^2 A \psi}{6} \left(1-5N+6N^2 + 2X(1-6N+3N^2)\right) \nonumber \\
&-&\frac{X^2 U_0 h^2 A \psi}{6}\left(X^2(5-7N)+2X^3\right)+ \ldots
\end{eqnarray}

One should note that $\epsilon$ is actually a decreasing function for all $y$. Moreover one can only trust the potential in the region
where $ \psi \sim 0$ since we have explicitly assumed that $x_i \sim L/2$ in the definition of the superfields. One can see that near the origin the
slow roll parameter has the form
\begin{equation}
\epsilon_X \sim \frac{2 \tilde A^2}{9} \frac{(1+X)^2(1-3N+2X)^2}{(1-2N+X)^2}
\end{equation}
which reproduces the result of (\ref{eq:epsilon2}) in the appropriate limit.
One notes that the value of $\epsilon_X$ decreases with increasing $N$, for fixed $X$. If one also increases $X$, then one requires larger and larger
values of $N$ in order to keep the parameter suppressed for inflation. Since inflation requires $\epsilon_X < 1$ we see that this requires
larger and larger $N$ as we increase the level of the interaction. To illustrate this we sketch the behaviour of the $\eta, \epsilon$ parameters as
a function of $\psi$ for fixed $N$ in Figures 1 and 2. One notes that inflation must occur in the region where $\psi \sim 0$, and therefore as one includes
higher order terms in $X$, this condition becomes impossible to satisfy. Numerically we find the following lower bound on the number of branes
\begin{equation}
N > N_c + \tau (X-2)
\end{equation}
valid for $X>1$, where $\tau$ is a constant of $\mathcal{O}(1)$.
The numerical coefficient $N_c$ is sensitive to the volume of the compact manifold. What is interesting to note is that
the bound shifts by a constant factor $\tau$ as we increase the level of interaction. Since we are taking $\mathcal{V}_{OM}$ to be order unity, this
implies $\tau$ is order unity, although the precise value also depends on the volume.
For the canonical choices discussed previously, we find that
$N_c \sim 33$ and $\tau \sim 7$.
The value of $N_c$ increases as we increase the volume, thus for $\mathcal{V} \sim 100$ we have $N_c = 23$ whilst for $\mathcal{V} \sim 1000$
we see that $N_c$ doubles to $N_c=46$. If one refers back to the bound in (\ref{eq:Nbound}) we see that the number of branes is bounded from above
by causality arguments. Inflation will then only occur if the number of branes falls in between the relevant bound. For example we see that
(setting $\mathcal{V}_{OM} \sim 7$)
\begin{eqnarray}
46+10(X-2) < &N& < 283 \hspace{0.5cm} \mathcal{V} \sim 1000 \nonumber \\
23+6(X-2) <& N& < 28 \hspace{0.65cm} \mathcal{V} \sim 100.
\end{eqnarray}
Note that the latter bound cannot be satisfied for $X>2$. Therefore eternal inflation will only occur precisely when $X=2$ and $23 < N < 28$
which is a highly tuned solution.

If one takes the potential $U_X$ at face value, then is appears that there is a large region of parameter space where inflation
(although not of the simple power law kind) is possible, although eternal. Since we cannot parameterise the end of inflation in an obvious manner,
the question of reheating is an important one. The issue is that the effective theory has a geometric origin when viewed from the UV, namely the
fivebranes must eventually collide with the boundary branes. If the inflationary phase lasts this long, then one assumes that reheating will occur
when the fivebranes merge with the boundary via instanton transitions. However this requires a more detailed understanding of the effective theory, since
other terms in the superpotential must be included.

In the absence of such a procedure, one may ask if instant preheating may be used both as an end for inflation and also as a means to reheat the
domain wall branes. A simple way in which this may be achieved is to consider an additional (effective) coupling between the inflaton and
the volume modulus via \cite{Felder:1998vq, Felder:1999pv, Panda:2009ji}
\begin{equation}
\mathcal{L} \sim  -\frac{1}{2} g_1^2 \psi_{*}^2 \tilde{s}^2 - g_2 \bar{\chi}\chi \tilde{s}
\end{equation}
where $\tilde{s}$ is related to the real part of the superfield through $\tilde{s}=M_p s$, $\chi$ is a bulk fermion field and $\psi_{*}$
represents the shifted inflaton $\psi_{*} = \psi-\psi_e$
such that inflation ends at $\psi_{*}=0$. Since we cannot analytically control the end of inflation in these models, we should take it
to represent the value of $\psi$ at which we expect other terms in the superpotential to become important.

\begin{figure}\label{figure:fig1}
\includegraphics[width=0.5\textwidth]{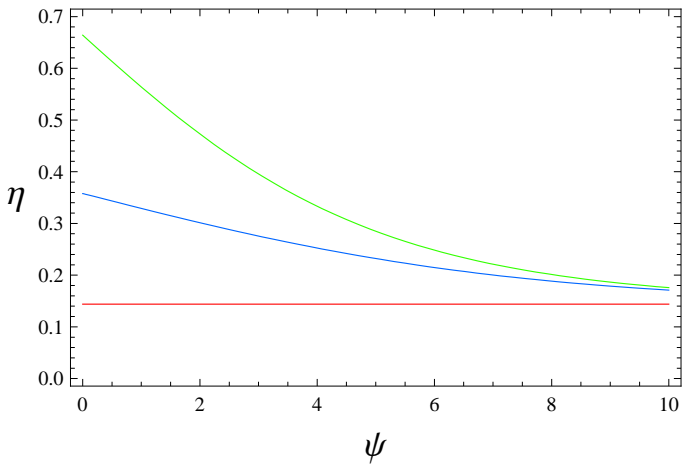}
\caption{Plot of $\eta(\psi)$ for $N=50$, using standard values for the volumes. The bottom (red) line corresponds to $X=1$ ie NN interaction.
The middle (blue) line corresponds to $X=2$, and the top line is $X=3$.}
\end{figure}

\begin{figure}\label{figure:fig2}
\includegraphics[width=0.5\textwidth]{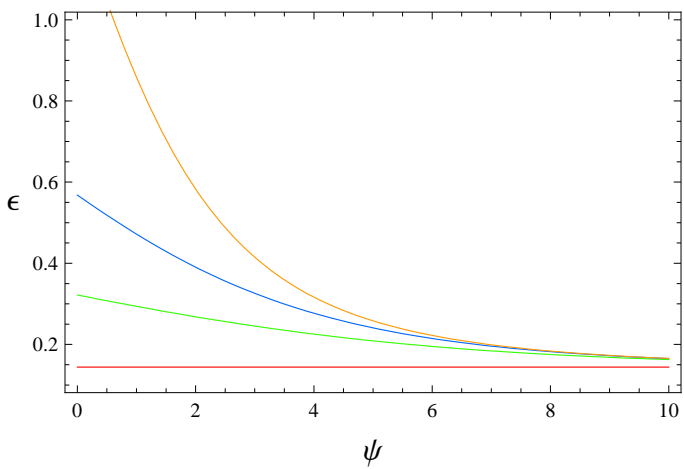}
\caption{Plot of $\epsilon(\psi)$ for $N=50$, using standard values for the volumes. The bottom (red) line corresponds to $X=1$ ie NN interaction.
The green line above this corresponds to $X=2$, the blue line is $X=3$ and the top line is $X=5$.}
\end{figure}

\section{Discussion}
\label{discussion}
In this paper we have re-considered the power law inflation scenario proposed in the Heterotic M-theory model of \cite{Becker:2005sg} to better
understand the regime of validity of the theory. As a consequence, we have determined the correct expression for the scalar potential in the presence of instanton
interactions between the $X$ nearest neighbours and shown that this also leads to eternal inflation. By requiring the solution to be $i)$ causal and $ii)$ have
inflating trajectories we find the following numeric bound on the number of branes, valid for $X>1$
\begin{equation}
N_c + \tau(X-2) < N < \frac{2 \mathcal{V}}{\mathcal{V}_{OM}}
\end{equation}
where both $N_c$ and $\tau$ depend on the two volumes although for a
large range of phenomenologically favoured parameters we expect $N_c
\sim \mathcal{O}(10), \tau \sim \mathcal{O}(1)$. The dependence on
the overall volume is the most sensitive, and we argued that for the
volume $\mathcal{V} \sim 100$, the above bound can only be satisfied
when $X=2$. This corroborates an intuitive result, namely that as we
increase the number of interactions $X$, the number of branes
required to meet our inflationary requirement must also increase.
This does not affect the $X=1$ solution
{\cite{Becker:2005sg}}, which
corresponds to power law inflation, precisely because of the special
nature of that model. More generally one can see that inflation with
an arbitrary number of branes, and an arbitrary number of instanton
interactions, is less likely. Whilst this is intuitive, it is worth
investigating in some detail precisely because the Heterotic theory
provides a very useful example where the standard model sector can
be identified (in principle).

Our results also suggest that despite the beautiful simplicity of the assisted inflation scenario, there is a large drawback in that the theory
breaks down once we begin to probe away from the $\psi \sim 0$ region due to the assumption that the branes are localised near the centre of the orbifold. Whilst
it is interesting to speculate on the small instanton transition which results in the branes dissolving into the boundary, one must bear in mind that the
effective theory is no longer valid in this region, and one must include terms arising at the same order in the superpotential.

\section*{Acknowledgments}
Thanks to Steve Thomas for his comments. JW is supported by EPSRC of Canada. PVM  research was
supported by the grant FEDER/POCI/FIS/P/57547/2004.

\appendix
\section{Phase space}
Since we have a canonical action, the equation of motion for the inflaton is given by the usual expression $\ddot{\psi}+3H\dot{\psi}+V'=0$. If we define
the following dimensionless variables
\begin{equation}
\Gamma = \frac{\psi}{M_p} \hspace{0.5cm} \Upsilon = \frac{\dot{\psi}}{M_p^2}
\end{equation}
then we can define the following autonomous equations for the flow of these parameters via
\begin{eqnarray}
\dot{\Gamma} &=& \Upsilon M_p \\
\dot{\Upsilon} &=& -\frac{V'}{M_p^2} \mp \sqrt{3}\Upsilon \left(\frac{V}{M_p^2}+ \frac{\Upsilon^2 M_p^2}{2} \right)^{1/2}
\end{eqnarray}
where primes denote derivatives with respect to $\psi$ and we will set $M_p=1$ at the end. The linearised perturbations about such a solution
are therefore
\begin{eqnarray}
\dot{\delta \gamma} &=& M_p \delta \upsilon \\
\dot{\delta \upsilon} &=& - \frac{\delta \gamma V_{\Gamma \Gamma}(\Gamma_c)}{M_p^3} \mp \sqrt{3} \delta \upsilon \sqrt{\frac{V(\Gamma_c)}{M_p^2}+\frac{M_p^2
\Upsilon_c}{2}} \nonumber \\
&\mp& \frac{\sqrt{3} \Upsilon_c}{2} \left(\frac{V(\Gamma_c)}{M_p^2}+\frac{M_p^2 \Upsilon_c}{2} \right)^{-1/2} \left(\frac{\delta x V_{\Gamma}(\Gamma_c)}{M_p^2}+
M_p^2 \delta \upsilon \right) \nonumber
\end{eqnarray}
where the ($\pm$) ambiguity arises from taking the square root of the Hubble equation, and $V_{\Gamma}(\Gamma_c)$ denotes a derivative with respect to $\Gamma$
and evaluated at $\Gamma=\Gamma_c$. This is a matrix equation where for stability of the fixed points we
require both eigenvalues to be negative (ie a positive determinant). To keep track of the $(\pm)$ ambiguity let us introduce $p=\pm$ and therefore the
eigenvalues become
\begin{eqnarray}
\lambda_{\pm} &=& \frac{p}{4\alpha}\left(-\sqrt{3}(2\alpha^2 + \Upsilon_c) \pm \sqrt{ \chi} \right)\\
\chi &=& 12\alpha^4-16\alpha^2V_{\Gamma \Gamma}+12\alpha^2 \Upsilon_c-8p\sqrt{3}\alpha^3 V_{\Gamma} \Upsilon_c + 3 \Upsilon_c^2  \nonumber
\end{eqnarray}
where we have used $V_{\Gamma \Gamma}=V_{\Gamma \Gamma}(\Gamma_c)$ and also
\begin{equation}
\alpha = \sqrt{\frac{V_{\Gamma \Gamma}(\Gamma_c)}{M_p^2}+\frac{M_p^2 \Upsilon_c}{2}}.
\end{equation}
Clearly the stability depends on the choice of $p$ and also the scalar potential.

One can study the parametric flow of $\Gamma$ and $\Upsilon$ as functions of time (assuming fixed $N,s,t,X$ and $M_p=1$), however the result is not illuminating
enough to show. What we see is that the fixed points of the system are localised at $\psi=0, \infty$ and that increasing $X$ forces the trajectory lines
to diverge from the $\psi=0$ point more and more rapidly. All the trajectories map onto one another as they approach the late time attractor point, as expected.

Let us consider the static solution $\Upsilon_c=0$ for simplicity,
since this is the only solution to $\dot{\Gamma}=0$. The other fixed
point therefore occurs when $V_{\Gamma} = 0$, or when the potential
is at an extremum. Since the potential is essentially runaway, this
condition can only be satisfied {for $\psi \to \infty$.
In this instance the} eigenvalue equation reduces to
\begin{equation}
\lambda_{\pm} = \frac{\sqrt{3}p \sqrt{V(\Gamma_c)}}{2M_p}\left(-1 \pm \sqrt{1-\frac{4 M_p^2 V_{\Gamma \Gamma}(\Gamma_c)}{3V(\Gamma_c)}}\right)
\end{equation}
and therefore we have the following solutions. For $p>0$ we find
$V_{\Gamma \Gamma}(\Gamma_c)/V(\Gamma_c)>0$ for stability and for $p
< 0$ we have $V_{\Gamma \Gamma}(\Gamma_c)/V(\Gamma_c)<0$. Ultimately
however the phase space dynamics tell us that
{this} fixed point solution occurs
when $\psi$ has rolled down to the bottom of the potential. However
this is beyond the regime of validity of the effective theory, since
other terms in the inflaton potential will be non-vanishing in this
regime.
%
%
%


\end{document}